# Effects of high vs moderate-intensity training on neuroplasticity and functional recovery after focal ischemia

Caroline PIN-BARRE, PhD[1]; Annabelle CONSTANS, MS[2]; Jeanick BRISSWALTER, PhD[1]; Christophe PELLEGRINO, PhD[3]; Jérôme LAURIN, PhD[2]

[1] Univ Nice Sophia Antipolis, Université de Toulon, LAMHESS, Nice, France

[2] Aix Marseille Univ, CNRS, ISM, Marseille, France

[3] Aix Marseille Univ, INSERM, INMED, Marseille, France

**Corresponding author:**
Dr. Jérôme LAURIN
Aix Marseille Univ
UMR CNRS 7287 « Institut des Sciences du Mouvement »
163, avenue de Luminy – CP 910
13288 Marseille cedex 09 - France
E-mail: jerome.laurin@univ-amu.fr
Tel.: +33 (0)4-91-82-84-11

**Cover title:** Effectiveness of endurance programs after stroke

**List of figures:**
**Figure 1:** Functional tests
**Figure 2:** Resting lactatemia, $S_{LT}$ and $S_{max}$
**Figure 3:** microglia and cytokines
**Figure 4:** $p75^{NTR}$
**Figure 5:** NKCC1/KCC2 ratio

**Number of words: 5958**

# Abstract

**Background and Purpose-**This study was designed to compare the effects of high-intensity interval training (HIT) and moderate-intensity continuous training (MOD) on functional recovery and cerebral plasticity during the first 2 weeks following cerebral ischemia.

**Methods-**Rats were randomized as follows: Control (n=15), SHAM (n=9), MCAO (n=13), MCAO-D1 (n=7), MOD (n=13) and HIT (n=13). Incremental tests were performed at day 1 (D1) and 14 (D14) to identify the running speed associated with the lactate threshold ($S_{LT}$) and the maximal speed ($S_{max}$). Functional tests were performed at D1, D7 and D14. Microglia form, cytokines, $p75^{NTR}$, KCC2 and NKCC1 expression were made at D15.

**Results-**HIT was more effective to improve the endurance performance than MOD and induced a fast recovery of the impaired forelimb grip force. The Iba-1 positive cells with amoeboid form and the pro- and anti-inflammatory cytokine expression were lower in HIT group, mainly in the ipsilesional hemisphere. A $p75^{NTR}$ overexpression is observed on the ipsilesional side together with a restored NKCC1/KCC2 ratio on the contralesional side.

**Conclusions-**Low-volume HIT based on lactate threshold appears to be more effective after cerebral ischemia than work-matched MOD to improve aerobic fitness, grip strength and might promote cerebral plasticity.

## Introduction

Ischemic stroke remains the leading cause of long-term physical disorders. Post-stroke hemiparesis frequently leads to physical deconditioning that strongly reduces the quality of life and represents an important burden on the family and society. Growing evidence from animal and human experiments indicated that aerobic training induced beneficial effects at the cardiovascular, muscular, cerebral and functional levels following cerebral ischemia[1,2]. Moderate-intensity aerobic training (MOD; for recommendations see[1]) is advised after stroke to improve the locomotor abilities, the peak oxygen uptake ($VO_{2peak}$) and the maximal running speed ($S_{max}$), which are strong indicators of quality of life. Early treadmill training in rodents could also promote functional recovery and cerebral plasticity by up-regulating the neurotrophin levels, enhancing synaptogenesis and limiting microglia-mediated pro-inflammatory cytokine release in the perilesional zones[3,4].

However, beneficial effects of MOD on functional recovery, aerobic fitness and quality of life remain frequently insufficient and controversial[1,5]. It is thus crucial to reconsider the current guidelines for exercise by defining a safe/effective dosage of training[5]. In this regard, authors recently showed that higher training intensities appeared promising for stroke patients[6]. Indeed, high-intensity interval training (HIT), known to be feasible and safe in moderate stroke patients[7], could improve $VO_{2peak}$, running economy and functional recovery but it remains controversial[6,7]. No clear evidence indicated whether the HIT effectiveness is more efficient on aerobic fitness and neuroplasticity than MOD[6,8]. Given that HIT is a time-efficient strategy, we postulated that it might accentuate functional recovery in the acute phase of cerebral ischemia compared to MOD.

In light of these considerations, the present study was designed to compare the effects of work-matched HIT and MOD programs on functional outcomes and cerebral plasticity during the first 2 weeks following cerebral ischemia in rats. One of the key points of the endurance protocols relies on determining for each animal the training intensity from an underestimated submaximal physiological parameter, i.e. the running speed associated with the lactate threshold ($S_{LT}$), which is relevant to distinguish high from moderate running speeds[9,10] and is highly sensitive to assess aerobic fitness[10,11]. In addition, the training effects on brain inflammation through microglia activation form was measured as well as the related expression of pro- (IL-1β, IL-12p40) and anti-inflammatory (IL-10) cytokines to determine the microglia function, which could be related to neurotrophin actions and synaptic plasticity[9]. Therefore, the pan-neurotrophin receptor p75 ($p75^{NTR}$) expression, known to strongly influence the neurotrophin functions after cerebral ischemia[12] was also assessed. The training-induced synaptic plasticity was observed through the expression of the potassium-chloride co-transporter (KCC2, a neuronal chloride extruder) and sodium-potassium-chloride co-transporter type 1 (NKCC1, an ubiquitously chloride importer) that are disturbed after cerebral ischemia and leads to alteration in the excitation/inhibition balance in brain[13].

## Material and methods

### *Animals*

Overall, 108 adult male Sprague-Dawley rats (250-270g; JANVIER®, France) were used but 70 of them were included. (please see http://stroke.ahajournals.org). Anesthesia and surgical procedures were performed according to the French law on animal care guidelines. Animal Care Committees of *Aix-Marseille Université* approved our protocol.

Each animal was randomly assigned to a group, making the impact of individuals less prominent: 1) Control (n=15), no surgery was performed that enabled to verify the reliability of measurements on 14 days; 2) SHAM (n=9), animals underwent surgery without cerebral ischemia to ensure that it did not affect measurements; 3) MCAO (n=13), animals underwent middle cerebral artery occlusion-reperfusion (MCAO-r) that enable to assess the spontaneous functional recovery and to verify the variance of rat activity level (no training); 4) MCAO-D1 (n=7) in which animals were sacrificed one day (D1) after MCAO-r to confirm that animals started training with a similar lesion severity; 5) HIT (n=13) and 6) MOD (n=13) in which animals underwent MCAO-r and carried out HIT and MOD programs respectively. (please see http://stroke.ahajournals.org).

### *MCAO-r and behavioral tests*

Rats were subjected to right MCAO-r for 2h. (please see http://stroke.ahajournals.org). The elevated body swing test (EBST), the ladder-climbing test and the forelimb grip force, were performed before (PRE) and after the surgery at day 1, 7 and 14 (D1, D7 and D14 respectively) following MCAO-r (**Figure I**). (please see http://stroke.ahajournals.org).

### *Incremental test*

Incremental tests were performed on 1° inclined treadmill at D1 and D14. These tests started with 5min of warm-up at 9m/min to reduce stress. Then, running speed was increased by 3m/min every 3min until animals could not maintain the imposed speed. $S_{max}$ was associated with the last reached stage. Each stage was separated by 20s interval to perform blood sampling (0.2μl) after partially cutting the distal area of the tail vein to determine $S_{LT}$. (please see http://stroke.ahajournals.org).

*Work-matched HIT and MOD programs on treadmill*

HIT and MOD programs included 10 sessions from D2 to D12 and two recovery days (D7 and D13) in order to reduce fatigue accumulation that may affect the incremental test performance. (please see http://stroke.ahajournals.org).

*Immunohistochemistry analysis*

Each animal was randomly assigned to either immunostaining analysis or Western blot at D15. Cresyl violet was used to measure the infarct volume and the percentage of tissue loss (% tissue loss). To investigate the changes of $p75^{NTR}$ and microglia form, immunostaining with antibodies against the $p75^{NTR}$ and Iba-1 were made at D15. (please see http://stroke.ahajournals.org).

*Western blot analysis*

To detect IL-10, IL-1β, IL-12p40, $p75^{NTR}$, KCC2 and NKCC1 expression, the total protein extracted from each frozen hemisphere was used for Western blot. (please see http://stroke.ahajournals.org).

*Statistical analysis*

Statistical analysis was performed using SigmaStat® software program (San Jose, CA, USA). All data are presented as Mean±SD. (please see http://stroke.ahajournals.org).

# Results

For the overall parameters, no difference was observed between Control and SHAM groups from PRE to D14. Likewise, no significant difference was observed at D1 for functional outcomes, infarct volume and endurance performance between MCAO, MCAO-D1, HIT and MOD groups indicating a similar lesion severity prior to training for each animal.

### *Endurance programs*

Running speed during MOD was lower than HIT during the 1$^{st}$ and 2$^{nd}$ training weeks (-28.9% and -31.2% respectively). Session duration of MOD group was higher than HIT group during the 1$^{st}$ and 2$^{nd}$ weeks (+49.2%; +59.8% respectively)**(Table I).**

### *Functional tests*

HIT induced a complete grip strength recovery without affecting other functional parameters. Indeed, grip force exerted by the affected forelimb decreased significantly between PRE and D1 for all injured-groups ($p<0.001$) and was significantly lower in MCAO, HIT and MOD groups than in Control and SHAM groups ($p<0.001$). Grip force remained significantly decreased ($p<0.01$) at D7 and D14 for both MCAO and MOD groups while it recovered in HIT group at D7 and D14 ($p<0.001$). Moreover, no difference was observed between Control, SHAM and HIT groups from D7 to D14 contrary to MCAO ($p<0.05$) and MOD ($p<0.001$) groups **(Figure 1.A)**. No difference was observed for both forelimbs force and for the non-affected forelimb force between groups from PRE to D14 (data not shown). The A/N ratio significantly decreased at D1 compared to PRE for MCAO, HIT and MOD groups ($p<0.001$) and was lower within all injured-groups than Control and SHAM groups ($p<0.001$). A/N ratio completely recovered only for HIT group from D7 to D14 compared to PRE ($p<0.001$). Likewise, A/N ratio of HIT group was significantly higher than MCAO and MOD groups from D7 to D14 (p<0.01) and remained similar to Control and SHAM groups, contrary to MCAO and MOD groups (p<0.01) **(Figure 1.B)**.

The left swings/total swings ratio (EBST) significantly increased for MCAO, HIT and MOD groups at D1, D7 and D14 compared to PRE (p<0.001) **(Figure 1.C)**.

The successful score (Ladder-climbing test) significantly decreased for MCAO, HIT and MOD groups at D1, D7 and D14 compared to PRE (p<0.001) without difference between groups. Nevertheless, this score was significantly higher for MCAO, HIT and MOD groups at D1 compared to D7 (p<0.001) and to D14 (p<0.001; p<0.01 and p<0.001 respectively) **(Figure 1.D)**.

*Incremental test*

HIT appeared to be more effective to recover aerobic fitness than MOD as indicated by changes in $S_{max}$ and $S_{LT}$. The resting blood lactate concentration of MCAO group at D14 (4.4±1.4mmol/L) was higher (p<0.001) compared to Control (2.2±0.6mmol/L), SHAM (2.0±0.6mmol/L), MOD (3.1±1.3mmol/L) and HIT (2.7±1.1mmol/L) groups **(Figure 2A)**.

$S_{LT}$ of MCAO, HIT and MOD groups at D1 were significantly lower than the one in Control (p<0.001) and SHAM (p<0.001) groups **(Figure 2.B)**. $S_{LT}$ significantly increased from D1 to D14 for HIT (20.4±2.4m/min for D1 and 34.5±3.8m/min for D14; p<0.001; **Figure 2.C**) and MOD (21.8±3.3m/min for D1 and 26.7±5.3m/min for D14; p<0.01) groups contrary to MCAO (22.5±3.9m/min for D1 and 22.5±4.5m/min for D14), Control (32.8±3.9m/min for D1 and 30.8±4.1m/min for D14) and SHAM (34.7±4.3m/min for D1 and 33.7±3.6m/min for D14) groups **(Figure 2.D)**. However, the $S_{LT}$ of HIT group at D14 was higher than MOD, MCAO groups (p<0.001). The $S_{LT}$ of MOD group was higher than $S_{LT}$ of MCAO group at D14 (p<0.01) but remained significantly lower than Control and SHAM groups (p<0.01), contrary to HIT.

$S_{max}$ at D1 was significantly lower in MCAO, HIT and MOD groups than Control and SHAM groups (p<0.001) **(Figure 2.B)**. However, $S_{max}$ significantly increased from D1 to D14 for HIT (26.1±3.5m/min for D1 and 40.8±5.9m/min for D14; p<0.001) and MOD (27.0±3.3m/min for D1 and 35.7±4.5m/min for D14; p<0.001) groups contrary to MCAO (27.5±4.9m/min for D1 and 30.0±5.4m/min for D14), Control (40.0±4.5m/min for D1 and 39.8±4.3m/min for D14) and SHAM (43.2±4.5m/min for D1 and 41.3±2.9m/min for D14) groups **(Figure 2.D)**. Moreover, the $S_{max}$ at D14 of HIT group was significantly higher than MOD (p<0.05) and MCAO (p<0.001) groups. The $S_{max}$ of MOD group was higher than $S_{max}$ of MCAO group at D14 (p<0.01) but remained significantly lower than Control and SHAM groups (p<0.05).

*Immunohistochemistry*

HIT promoted ramified microglia, p75 increase, restoration of NKCC1/KCC2 ratio and down-regulated pro- and anti-inflammatory cytokine expression, without affecting infarct volume and the percentage of tissue loss. Indeed, the number of amoeboid Iba-1+ cells for HIT group (20.6%±5.2) was significantly lower than MOD (74.9%±29; p<0.01) and MCAO (77.1 %±31.5; p<0.001) groups within the peri-lesional site as well as in the contralesional hemisphere (HIT: 11.3%±3.1; MOD: 54.3%±22.2; p<0.01; and MCAO: 46.9 %±29.9; p<0.01)**(Figure 3)**.

For qualitative staining, the cells of damaged hemispheres expressed $p75^{NTR}$ proteins in all lesioned groups contrary to SHAM group **(Figure 4.C)**.

No difference was observed between lesioned groups for infarct size and percentage of tissue loss (MCAO: -3.3±7.6%; HIT: -5.4±6.1%; MOD: -2.7±7.8% of the contralesional hemisphere).

*Western Blotting*

In the HIT group, IL-10 expression was significantly down-regulated in the ipsilesional hemisphere when normalized to the IL-10 expression of MCAO group (0.75±0.09; $p<0.01$) contrary to MOD (1.03±0.23). IL-12p40 expression was significantly down-regulated in the ipsilesional hemisphere after HIT (0.81±0.03, $p<0.01$). No difference was observed for IL-12p40 expression in MOD group (0.93±0.15). Likewise, no difference was observed for IL-1β between groups (0.85±0.23 for HIT and 0.93±0.12 for MOD).

The relative expression of $p75^{NTR}$ protein within the ipsilesional hemisphere in HIT (4.4±2.7; $p<0.01$) and MOD (3.1±1.9; $p<0.05$) groups was significantly higher than SHAM, contrary to MCAO (2.8±2.6) **(Figure 4.A)**. In the contralesional hemisphere, the $p75^{NTR}$ expression was not different between groups **(Figure 4.B)**.

In the ipsilesional hemisphere, no difference was observed for NKCC1/KCC2 ratio between MCAO (2.20±1.46), MOD (4.23±5.16), HIT (1.75±0.89) and SHAM (1.05±0.68) *(data not shown)*. The NKCC1/KCC2 ratio of HIT group (0.62±0.16) was significantly lower than MCAO (1.16±1.36; $p<0.05$) and MOD (2.42±1.38; $p<0.05$) groups in the contralesional hemisphere **(Figure 5)**.

At D1 and D15, grip force of the left forepaw, infarct volume and immunohistochemistry results were not correlated within groups *(data not shown)*.

## Discussion

For the first time, this study demonstrated that 2 weeks of HIT was more effective than a work-matched MOD program on multi-scale measurements after cerebral ischemia. The use of lactate threshold enabled defining high (>25m/min) and moderate (<20m/min) running speed in an individualized manner for each animal with cerebral ischemia. Such intensity ranges were not in accordance to previous studies in which exercise intensity between 10-13m/min was considered as intense for MCAO rats[10]. The difference might be explained by their use of empirical training intensities or maximal parameters ($S_{max}$ or $VO_{2peak}$), that were not highly relevant to distinguish high from moderate intensities[9–11]. Indeed, the ability to prescribe the optimal training stimulus might be greater whether intensity was based on submaximal physiological parameter that could be reached by the majority of patients, contrary to maximal parameters[11,14]. In our study, rats could begin an early individualized intense program that seemed not to be deleterious for functional recovery and infarct volume[15]. It was in accordance to previous studies showing that treadmill training starting during the 5 first days induced beneficial effects on recovery (contrary to training began within 24h post-ischemia in both rodent[2] and human[16]). It was also found in human that very early constraint-induced movement resulted in less motor improvement at 90 days[17]. However, our results did not indicate that HIT need to be performed in moderate stroke patients during the first two weeks because the initiation of aerobic program should be later to be feasible and safe (during the subacute phase, i.e. the first 3 months)[18,19]. Moreover, HIT induced rapid physiological adaptations although its session duration was shorter than work-matched MOD confirming that HIT is time-efficient[20]. It was important given that 'lack of time' remains a major barrier for patients to exercise participation. Interestingly, HIT could also elicit higher enjoyment than MOD despite higher ratings of perceived exertion during intense series. After reporting an initial apprehension, the patient confidence progressively increased during HIT[6]. It suggested that higher intensities might not be considered as a major barrier for patients.

Our study revealed a maintained decrease of $S_{LT}$ and $S_{max}$ during the first 14 days after MCAO-r for non-trained animals. It was thus strongly argued that spontaneous aerobic fitness recovery was insufficient. It was in accordance with another study in which a decrease of $S_{LT}$ was observed two days post-ischemia[21]. The decreased $S_{LT}$ at D1 might be associated with sensorimotor alterations such as interlimb coordination or strength deficits given that neuromuscular disorders might disturb metabolic activity. Given that the $S_{LT}$ was influenced by muscular typology composition and atrophy (observed from D7 after cerebral ischemia[10]), muscle typology changes might partially explain the decrease of $S_{LT}$ at D14, but not at D1.

HIT was more effective than MOD to improve aerobic fitness as indicated by a superior shift of $S_{LT}$ and $S_{max}$ to higher intensities during incremental test. Other studies indicated that HIT induced higher ventilatory threshold improvements than MOD in cardiovascular patients[22]. It thus suggested that rats were able to exercise for longer durations at greater percentages of their $S_{max}$, reducing fatigue at a given intensity after HIT. Both programs are known to improve the maximal oxidative capacity in humans and animals by increasing $VO_{2peak}$, contributing to explain $S_{max}$ improvement[15]. However, HIT further improved maximal running performance in our study as it was recently observed in rats with chronic heart failure[23].

Our study also revealed a resting hyperlactatemia 14 days after the cerebral ischemia in non-trained rats. High lactate levels during the acute phase of stroke (< 3 months) was also observed on 25% of stroke patients and seemed to have deleterious repercussions on functional recovery[24]. It might be suggested with caution that increase of cortisol and catecholamine blood levels induced by cerebral ischemia are known to be involved in blood lactate concentration accumulation, but also, that anaerobic glycolysis promoted by cerebral hypoperfusion in affected neurons might facilitate an increase in blood lactate[25]. Given that HIT and MOD are known to increase the lactate transporter expression, the lactate clearance might be improved by training preventing hyperlactatemia.

The decrease of affected forelimb grip force and its consequent strength asymmetry persisted until D14. However, HIT induced a rapid recovery of the affected forelimb grip force from D7 without acting on the other behavioral tests. The force improvement might be partially explained by a facilitation of fast motor units recruitment during HIT sessions because of running speed was above the lactate threshold (contrary to MOD). The force improvement of forelimb flexor muscles might be possible because they were activated during locomotion (and not only extensor muscles) that might promote muscular changes. Given that only grip force was improved, HIT might be combined with skilled reaching training, known to improve limb function after stroke, in order to maximize recovery.

However, no functional improvements were observed after MOD, similarly to previous reports[26]. In humans, beneficial effects of MOD on motor functions and quality of life were not consistently observed[18]. In rodents, treadmill training did not have a significant effect on limb function[2]. It was also demonstrated that running at a very low intensity (10.5m/min) was not sufficient to observe a recovery[27]. We thus postulated that the intensity for MOD (-20% of $S_{LT}$) was too low to observe benefits.

The amoeboid Iba-1+ cells, known to secrete pro-inflammatory agents and free radicals, was higher in MCAO and MOD animals within both hemispheres (which was not always observed following MCAO-r[28]) than in HIT animals. It reflected an inflammatory state, even after MOD. We postulated that MOD exercise intensity might not be sufficient to promote significant changes in microglia morphology, indicating that running intensity under the LT was unlikely the most effective strategy to reduce inflammation. Conversely, the Iba-1+ cells after HIT mainly showed ramified form within both hemispheres. This was in accordance with previous studies showing that physical training might induce microglia morphological changes[29]. However, the ramified form could exert either detrimental activity (M1 phenotype), characterized by pro-inflammatory cytokine release, or beneficial activity (M2 phenotype) by secreting anti-inflammatory cytokines and neurotrophins. In the present study, HIT, but not MOD, might decrease neuroinflammation in the ipsilesional hemisphere by down-regulating both pro- and anti-inflammatory cytokine expression. It was in accordance with findings revealing that aerobic training might be beneficial for neuroprotection by reducing the pro-inflammatory cytokine expression in healthy animals[30] and in mice with neurodegenerative diseases[29]. However, our results disagreed with a study on healthy mice in which vigorous training on treadmill enhanced anti-inflammatory cytokine IL-10 release[30]. The different training protocols and rodent models (healthy vs. cerebral ischemia) might explain controversial findings. Alternatively, it might also be possible that microglia progressively returned to resting state after HIT, resulting in a down-regulation of IL-10 level at D15. Therefore, the time course of cytokines during aerobic training remains to be investigated. Nevertheless, the absence of an IL-10 up-regulation after MOD was consistent with the observed amoeboid Iba-1+ cells, which was in accordance with a previous study[30]. Finally, results should be interpreted with caution because of the sample size for immunohistochemistry analysis of injured groups. However, we observed similar outcomes in each group without excessive variability that might be due to the individualization of training programs and to a strict control of behavioral outcomes after MCAO-r. Several studies, focused on brain structures with larger sample size, are required to clarify the effects of exercise intensity on cytokine expression and microglia function after cerebral ischemia. It might be possible that a decrease of a pro-inflammatory state promoted an adaptive neuroplasticity even if the influence of neuroinflammation on neuroplasticity remains to be explored[31].

To determine if HIT or MOD might influence brain plasticity, we quantified the p75$^{NTR}$ level, which strongly contribute to mediate the neurotrophin cellular functions during embryonic development and following central nervous system (CNS) lesion[12]. Our study was the first to measure the effectiveness of different training intensities on p75$^{NTR}$ expression within both hemispheres after severe cerebral ischemia. We found that HIT induced an increase of p75$^{NTR}$ at ipsilesional level. However, p75$^{NTR}$ expression could be associated with beneficial or detrimental functions complicating result interpretation. It was thus difficult to establish whether the increase of p75$^{NTR}$ expression was associated with cellular death processes or with beneficial neuroplasticity. Nevertheless, several results allowed us to suggest that the p75$^{NTR}$ expression might be beneficial. First, aerobic training stimulated endogenous BDNF/TrkB expression (BDNF, brain-derived neurotrophin factors) within both hemispheres that is known to influence p75$^{NTR}$ expression[4,32]. Moreover, BDNF suppression, by injecting TrkB-FC, inhibited the increase of p75$^{NTR}$ expression after axotomia reducing the neuronal survival[33]. Only one study found on aged rats that, following 8 weeks of endurance training, the p75$^{NTR}$ level increased in parallel with an enhancement of BDNF expression[34]. Authors postulated that p75$^{NTR}$ increase might promote survival of damaged neurons, trigger apoptosis for cleaning debris and induce beneficial environment for axonal regrowth and inflammatory prevention[12]. In addition, it was demonstrated in healthy rats that high-intensity training could induce higher cerebral concentrations of BDNF and GDNF compared to MOD, which was related to neuroprotection[4,35]. To reinforce our hypothesis on the p75$^{NTR}$ role, it was demonstrated that aerobic training could facilitate the conversion of the proBDNF to the mature BDNF (mBDNF) in the peri-ischemic regions, which was associated with functional improvements[4,29,36]. Indeed, the decrease of proBDNF expression was associated with lower cell death and synaptic depression. On the other hand, the increase of mBDNF expression could promote synaptic plasticity and rescue neuronal loss[37]. In light of these findings, we suggested that the increase of p75$^{NTR}$ expression after HIT might reflect a beneficial role of neurotrophin expression. Moreover, circulating BDNF levels was not measured because it does not mirror brain BDNF levels after stroke. It thus complicates interpretation and might be less relevant than brain BDNF.

Finally, in order to link neurotrophin action to brain plasticity, we studied the Cl$^-$ homeostasis through the KCC2 and NKCC1 expression, proteins known to be sensitive to neurotrophin levels. Indeed, BDNF could promote KCC2 expression after CNS trauma. To our knowledge, no study determined the role of different aerobic trainings on the Cl$^-$ co-transporters following cerebral ischemia despite their crucial role in the CNS function[38]. Only one study indicated that these chloride co-transporters were sensitive to aerobic training because such physical activity could affect the spinal KCC2 and NKCC1 expression after spinal cord injury, in parallel with functional recovery improvement[39]. We thus postulated that HIT could influence the Cl$^-$ homeostasis by changing the KCC2 and NKCC1 expression after cerebral ischemia that might optimize the equilibrium between excitation/inhibition in brain cells. It also appeared interesting to postulate that the contralesional hemisphere was sensitive to brain plasticity, as indicated by the decrease of NKCC1/KCC2 ratio after HIT, together with changes in reactive gliosis and inflammation[40].

## Conclusion

This study provided new promising insights into the effectiveness of low-volume HIT on the physiological determinants of aerobic fitness and grip strength. It also seemed that HIT might promote neurotrophin action, synaptic plasticity compared with work-matched MOD. According to human studies, results needed to be interpreted with caution because risk factors and comorbidities were not taken into account in the present study that might change the effects of these endurance programs. This study needed to be considered as an initial assessment of the effects of these training protocols. Nonetheless, HIT is known to be feasible in moderate stroke patients but its effectiveness compared to MOD needs to be assessed. In animals, it was recommended to deepen the neuroplasticity mechanisms induced by HIT without forgetting its outcomes on functional recovery. It seems now critical to bring evidence on the effects of detraining for the different aerobic programs that remains poorly investigated, but important for improving the long-term recovery.

## Acknowledgments

None

## Sources of Funding

None

## Disclosures

None

# Figure Legends

## Figure 1
*Forelimb grip force.* **A)** Affected forelimb, **B)** A/N ratio.
*significant decrease from D1 to D14 in injured-groups compared to PRE (except for HIT at D7 and D14). #significant increase from D7 to D14 for HIT group compared to PRE. +significant lower force in injured-groups compared to non-injured groups from D1 to D14 (except for HIT at D7 and D14). §significant higher force for HIT group compared to MOD and MCAO groups at D7 and D14.
**C)** *The elevated body swing test.*
*left swings/total swing increase at D1, D7 and D14 compared to PRE for injured-groups.
+higher left swings/total swing for MCAO, HIT and MOD groups compared to Control and SHAM groups from D1 to D14.
**D)** *Successful score* (% of PRE).
*significant decrease at D1, D7 and D14 compared to PRE for injured-groups. The successful score increased from D1 to D14 for MCAO (#), HIT (§) and MOD (φ).
+lower successful score for injured-groups compared to Control and SHAM groups.

## Figure 2
**A)** *Resting blood lactate.*
*resting lactatemia (mmol/L) increase between D1 and D14 for MCAO group ($p<0.001$).
+higher resting lactatemia for MCAO group compared to Control, SHAM, HIT and MOD groups at D14 ($p<0.001$).
**B)** $S_{LT}$ *and* $S_{max}$ *(m/min) at D1 after cerebral ischemia.*
+lower $S_{LT}$ and $S_{max}$ for MCAO, HIT and MOD groups compared to Control and SHAM ($p<0.001$).
**C)** $S_{LT}$ *and* $S_{max}$ *(m/min) after training.*
*$S_{LT}$ and $S_{max}$ increase from D1 to D14 for HIT and MOD groups ($p<0.001$ and $p<0.01$ respectively).
+higher $S_{LT}$ and $S_{max}$ at D14 for HIT, Control and SHAM groups compared to MCAO and MOD groups ($p<0.001$ for $S_{LT}$; $p<0.05$ for $S_{max}$ compared to MOD group).
φhigher $S_{LT}$ and $S_{max}$ at D14 for MOD group compared to MCAO group ($p<0.05$ for $S_{LT}$ and $p<0.05$ for $S_{max}$).
**D)** *Example of lactatemia kinetic (raw data) during incremental test before and after HIT.*
At the D14 incremental test, $S_{LT}$ was observed at a higher running speed (36m/min) compared to D1 (24m/min). Arrows indicate the lactate threshold.

## Figure 3
*The impact of exercise training on microglial cells morphology and cytokine expression.*
**A) Quantification of microglia morphology changes at peri-ischemic level in ipsilesional** (above figure) **and in contralesional** (bottom figure) **hemisphere at D15.** In both hemispheres, the % of amoeboid Iba-1+ cells is significantly lower in HIT group than in MCAO (+, $p<0.001$ and $p<0.01$) and MOD (*; $p<0.01$ and $p<0.01$) groups. **B) Example of immunofluorescent staining with Iba-1 protein on MCAO, MOD and HIT groups in ipsilesional and contralesional hemispheres.** Green arrows indicate ramified reactive microglial cells whereas red arrows indicate amoeboid microglial cells. IP: ipsilesional; CT: contralesional. **C) IL-10 expression of HIT and MOD groups normalized to MCAO group condition in the ipsilesional hemisphere at D15** (left figure). *IL-10 expression decrease of HIT group (p<0.01). Representative immunoblot of IL-10 protein (and α-tubulin) in the ipsilesional hemisphere of injured and SHAM groups (right figure). **D) IL-12p40 expression of HIT and MOD groups normalized to MCAO group condition in the ipsilesional hemisphere at D15** (left figure). *IL-12p40 expression decrease of HIT group (p<0.01). Representative immunoblot of IL-12p40 protein (and α-tubulin) in the ipsilesional hemisphere of injured and SHAM groups (right figure).

**Figure 4**
**The impact of exercise training on p75$^{NTR}$ expression.**
**Expression of p75$^{NTR}$ normalized to SHAM condition. A)** # indicates a significant increase of p75$^{NTR}$ expression in HIT (p<0.001) and MOD (p<0.01) groups compared to SHAM group in the ipsilesional hemisphere at D15. Representative immunoblot of p75$^{NTR}$ protein (and GAPDH) in the ipsilesional hemisphere of injured groups. **B)** No difference in the contralesional hemisphere. **C)** Illustration of p75$^{NTR}$ immunostaining in the ipsilesional hemisphere.

**Figure 5**
**The impact of exercise training on NKCC1/KCC2 ratio in the contralesional hemisphere.**
A) The NKCC1/KCC2 ratio is significantly lower in HIT group than the one of MOD (*) and MCAO ($^+$) groups. **B) Representative immunoblot of KCC2 and NKCC1 proteins (and GAPDH or β-tubulin).**